\documentclass[aps,twocolumn,prb,amsmath,amssymb]{revtex4}
\usepackage{multirow}
\usepackage{graphicx,epsfig}
\usepackage{wasysym}

\begin{document}
\title{Series expansion of the quantum admittance in mesoscopic systems}
\author{Adeline Cr\'epieux}
\affiliation{Aix-Marseille Universit\'e, CNRS, CPT UMR 7332, 13288, Marseille, France}

\begin{abstract}
The quantum admittance of an interacting/coupled mesoscopic system and its series expansion are obtained by using the refermionization method. With the help of these non-perturbative results, it is possible to study the dependencies of the admittance according to the applied dc voltage, temperature, and frequency without any restriction on the relative values of these variables. 
Explicit expressions of the admittance are derived both in the limits of weak and strong interactions/coupling strength, giving clear indication of the inductive or capacitive nature of the mesoscopic system. They help to determine the conditions under which the phase of the current with respect to the ac voltage is positive.
\end{abstract}

\maketitle

\section{Introduction}

There is a growing interest regarding the quantum admittance of mesoscopic systems from both experimental and theoretical points of view. The reason is that this quantity contains in itself information about the dynamics of these systems: its real part provides the value of low-frequency effective resistance and is related to the anti-symmetrized current fluctuations,\cite{bena07,safi08} whereas its imaginary part gives information about  the inductive or capacitive nature of the system. Moreover both its real and imaginary parts make it possible to determine the phase shift of the current response with respect to the applied voltage, and appear in the expression of non-symmetrized noise\cite{zamoum12} which is the relevant quantity in high frequency experiments.\cite{aguado00,deblock03,billangeon06,onac06} 
 
Measurement of quantum admittance has been achieved recently in two dimensional electron gas,\cite{gabelli06,gabelli07,gabelli12,hashisaka12} carbon nanotube double quantum dot,\cite{chorley12} superconducting junction\cite{basset12} and quantum dot coupled to a two dimensional electron gas.\cite{frey12}
A formalism to calculate the quantum admittance for non-interacting systems based on scattering theory is available in the literature.\cite{fu93,christien96a,christien96b,pretre96,buttiker96} Effect of strong electron interactions on the admittance of a quantum dot coupled to an edge state has also been considered,\cite{hamamoto10} but distinct approaches were used to describe the perturbative regime (weak barrier case) and the non-perturbative one. Admittance of a double quantum dot has been calculated with an emphasis on the effective capacitance obtained in the low frequency limit.\cite{cottet11}

\begin{figure}[!h]
\begin{center}
\includegraphics[width=9cm]{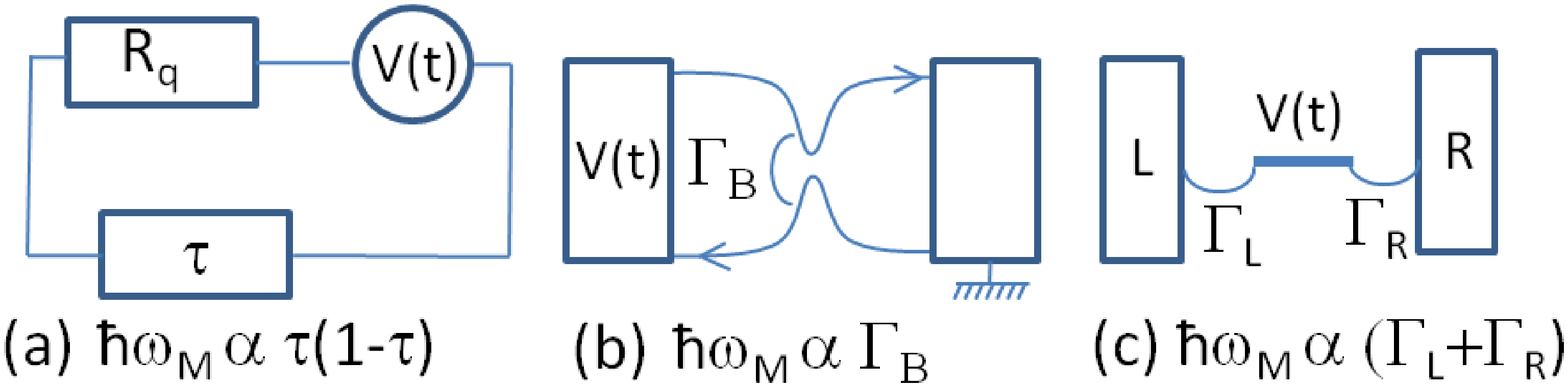}
\caption{Pictures of (a) a one-channel conductor with transmission $\tau$ coupled to a resistance equals the quantum of resistance  $R_q=h/e^2$, (b) a constriction in a fractional quantum Hall bar with backscattering amplitude $\Gamma_B$, and (c) a double barrier quantum dot whose coupling amplitudes with the left (L) and right (R) reservoirs are denoted $\Gamma_L$ and $\Gamma_R$, respectively. The associated characteristic energy, $\hbar\omega_M$, is indicated for each of these systems. The coefficient $\nu$ (see text) equals $1/2$ for systems (a) and (b), and $1$ for system (c).
\label{figure1}}
\end{center} 
\end{figure}

The present work focuses on the calculation of the quantum admittance using a refermionization method which makes it possible to describe entire frequency, temperature, and voltage ranges for various coupled or interacting mesoscopic systems (see Fig.~\ref{figure1}) in an unified way. Indeed, our results apply to: (i)~a one-channel conductor coupled to a measurement circuit the resistance of which equals the quantum of resistance, an experimental configuration which is realizable with a quantum point contact;\cite{parmentier11} (ii)~a constriction in a two dimensional electron gas in the fractional quantum Hall regime with a specific filling factor; and (iii)~a quantum dot coupled to two reservoirs.

The paper is organized as follows: In Sec.~II, the procedure for calculating the admittance is formulated and the model used to describe the systems of Fig.~\ref{figure1} is presented. The results are reported in Sec.~III. In Sec.~IV are emphasized the different regimes that are reached when voltage, temperature and frequency vary with respect to each other. The conclusion is given in Sec.~V. Technical details of calculations are described in the Appendixes.
 

\section{Formulation}

When an ac voltage superposed to a dc voltage $V(t)=V_0+V_\omega\cos(\omega t)$ is applied to an electronic circuit, the current becomes time dependent and can be written as a series (photo-assisted current): $I(t)=\sum_{N=-\infty}^{\infty}I^{(N)}(\omega)e^{iN\omega t}$, where $I^{(N)}$ is the $N^{th}$ harmonic of the current. By using the fact that it is a real quantity, the current can be written equivalently under the form:
\begin{eqnarray}
 I(t)=I^{(0)}+2\sum_{N=1}^{\infty}\mathrm{Re}\{I^{(N)}(\omega)e^{iN\omega t}\}~.
\end{eqnarray}
In the linear response regime with respect to the ac voltage, i.e. $e^*V_\omega\ll\{e^*V_0,\hbar\omega\}$ where $e^*=\nu e$ is the effective charge of the carrier that travels through the mesoscopic system, the current reduces to:\cite{note1}
\begin{eqnarray}\label{current}
I(t)\approx I^\mathrm{dc}(V_0)+2\nu V_\omega\mathrm{Re}\{Y(\omega)e^{i\omega t}\}~,
\end{eqnarray}
where $I^\mathrm{dc}$ is the dc current response to a constant voltage $V_0$, and $Y$ is the admittance defined as the derivative of the first harmonic of the current with respect to the amplitude of the ac voltage: $Y(\omega)=\nu^{-1}\partial I^{(1)}(\omega)/\partial V_\omega$. The use of the coefficient $\nu$ makes it possible to treat the three systems depicted in Fig.~\ref{figure1} in an unified way. From Eq.~(\ref{current}), one can see that the real part of the admittance (the conductance) gives the instantaneous response to the ac voltage, whereas the imaginary part of the admittance (the susceptance) is responsible for the phase difference between the current response and the ac voltage. Indeed, introducing the phase $\varphi(\omega)=\arctan[\mathrm{Im}\{Y(\omega)\}/\mathrm{Re}\{Y(\omega)\}]$, the current reads as: $I(t)=I^\mathrm{dc}(V_0)+2\nu|Y(\omega)|V_\omega\cos(\omega t+\varphi(\omega))$. The method used here to get the admittance consists of calculating the photo-assisted current, then extracting its first harmonic and finally taking its derivative with respect to $V_\omega$. Studies of photo-assisted current and electrical response to time-dependent voltage in interacting mesoscopic systems have been achieved by several authors\cite{sharma01,feldman03,schmidt07,ma11,safi11,dolcini12} but they did not notice the connections with the admittance.

The systems depicted in Fig.~\ref{figure1} are modeled in the framework of the Tomonaga-Luttinger theory,\cite{tomonaga50,luttinger63} by the Hamiltonian:\cite{note2}
\begin{eqnarray}~\label{hamiltonian}
H&=&\frac{\hbar v_F}{4\pi}\int_{-\infty}^{\infty}dx[(\partial_x\phi_-(x))^2+(\partial_x\phi_+(x))^2]\nonumber\\
&&+\frac{\hbar\omega_M}{4\pi}e^{i[\phi_-(x)+\phi_+(x)]/\sqrt{2}-ie^*\chi(t)/(\hbar c)}+hc~,
\end{eqnarray}
where $\phi_{-}$ and $\phi_{+}$ are the bosonic fields associated with the left~($-$) and right~($+$) moving electrons, and $v_F$ is the Fermi velocity. The function $\chi(t)=-c\int V(t)dt$ is included in order to treat the time-dependent applied voltage. The energy $\hbar\omega_M$ characterizes the mesoscopic system:~ (i) in the case where the mesoscopic system is coupled to a measurement circuit with a resistance equal to the quantum of resistance (see Fig.~\ref{figure1}(a)), then $\hbar\omega_M\propto (1-\tau)\tau$, where $\tau$ is the transmission through the mesoscopic system, and $\nu=1/2$ because the bias voltage seen by the mesoscopic system is $V(t)/2$;\cite{zamoum12} (ii) in the case of a constriction in a two dimensional electron gas in the fractional quantum Hall regime with a filling factor $\nu=1/2$  (see Fig.~\ref{figure1}(b)), then $\hbar\omega_M\propto\Gamma_B$ where $\Gamma_B$ is the backscattering amplitude;\cite{crepieux04} (iii) in the case of a quantum dot coupled to left (L) and right (R) reservoirs with amplitudes $\Gamma_L$ and $\Gamma_R$  (see Fig.~\ref{figure1}(c)), then $\hbar\omega_M\propto\Gamma_L+\Gamma_R$,\cite{fu93} and $\nu=1$. In the first two cases, the calculated current is the backscattering current whereas in the latter case, it corresponds to the flux of charges across the barriers.

The justification of Eq.~(\ref{hamiltonian}) is obvious for system~(b) of Fig.~\ref{figure1} since the transport of charge carriers in such a system takes place along the edge states of the Hall bar: due to their one-dimensional character the edge states are well described within the Tomonaga-Luttinger theory allowing one to treat the Coulomb interactions.\cite{note2} Concerning the system~(a) of Fig.~\ref{figure1}, the justification for using Eq.~(\ref{hamiltonian}) is based on the mapping that has been established between a one-channel conductor coupled to a measurement circuit and one impurity in a Luttinger liquid.\cite{safi04} Successful description of transport properties has been recently achieved by this means.\cite{zamoum12,parmentier11,jezouin13} Moreover, since the expression of the admittance obtained from Eq.~(\ref{hamiltonian}) is identical to the result obtained for system (c) of Fig.~\ref{figure1} (see Appendix \ref{appA} for a detailed calculation), it is possible to include the double barrier quantum dot in the list of interacting/coupled systems described here.


\section{Results}

From the Hamiltonian of Eq.~(\ref{hamiltonian}), the photo-assisted current can be calculated with the help of a refermionization procedure\cite{chamon96} which has the great advantage to be a non-perturbative method: it leads to exact results whatever the value of the energy $\hbar\omega_M$. The first order harmonic of the ac current in the linear response regime with $V_\omega$ has been derived in Ref.~\onlinecite{crepieux04}, it leads to the admittance:
\begin{eqnarray}\label{exp_admi}
Y(\omega)&=&\frac{e^2}{2h\omega}\int_{-\infty}^{\infty}\big[t(\Omega)-t(\Omega-\omega)\big] \nonumber\\
&&\times\big[f(\hbar\Omega+e^*V_0)-f(-\hbar\Omega+e^*V_0)\big]d\Omega~,
\end{eqnarray}
where $f$ is the Fermi-Dirac distribution function, and $t(\Omega)=(i\omega_M/2)/(\Omega+i\omega_M/2)$ is the transmission amplitude through: the one-channel conductor of Fig.~\ref{figure1}(a), the constriction of Fig.~\ref{figure1}(b), or the barriers of Fig.~\ref{figure1}(c). Note that the real part of the admittance obeys the relation\cite{tucker85,safi10,zamoum12} $\mathrm{Re}\{Y(\omega)\}=e\sum_\pm [\pm I^\mathrm{dc}(V_0\pm\hbar\omega/e^*)/(2\hbar\omega)]$.

To characterize its low frequency behavior, the admittance can be expanded in powers of $\omega$ according to $Y(\omega)=\sum_{n=0}^\infty Y^{(n)}\omega^n$, $Y^{(n)}$ is the $n$-order harmonics, with the help of the relation:
\begin{eqnarray}
t(\Omega-\omega)=\sum_{n=0}^\infty \left(\frac{2\omega}{i\omega_M}\right)^nt^{n+1}(\Omega)~.
\end{eqnarray}
The details of the calculation are given in Appendix \ref{appB}, we obtain:
\begin{eqnarray}\label{coeff_n}
Y^{(n)}&=&-\frac{2^{n}e^2}{h(i\omega_M)^{n+1}}\int_{-\infty}^{\infty}\left[t(\Omega)\right]^{n+2}\nonumber\\
&&\times\left[f(\hbar\Omega+e^*V_0)-f(-\hbar\Omega+e^*V_0)\right]d\Omega~,
\end{eqnarray}
which is purely real for odd values of $n$ and purely imaginary for even values of $n$. Eqs.~(\ref{exp_admi}) and (\ref{coeff_n}) are the central results of this work since they make it possible to calculate the admittance, and all the coefficients of its expansion in powers of $\omega$, whatever the dc voltage $V_0$, temperature $T$, and frequency $\omega_M$. In the next section, the behavior of the admittance in various regimes is discussed by examining the combined effects of the energies $\hbar\omega_M$, $e^*V_0$, $\hbar\omega$, and $k_BT$.


\section{Discussion}

\subsection{Zero temperature limit}

In this limit, the integration over frequency in Eq.~(\ref{exp_admi}) can be performed explicitly:
\begin{eqnarray}\label{Y_T0}
Y_{T=0}(\omega)&=&\frac{e^2}{4i h}\frac{\omega_M}{\omega}\sum_\pm\mathrm{ln}\left(1+\frac{i\hbar\omega}{\hbar\omega_M/2\pm ie^*V_0}\right)~.\nonumber\\
\end{eqnarray}
The real and imaginary parts of Eq.~(\ref{Y_T0}), which are related by the Kramers-Kronig relation, coincide with the one obtained with the help of scattering theory in the case of a double barrier quantum dot.\cite{fu93,buttiker96} However, the compact writing of the zero temperature admittance as formulated by Eq.~(\ref{Y_T0}) is novel.

\begin{figure}
\begin{center}
\includegraphics[width=4cm]{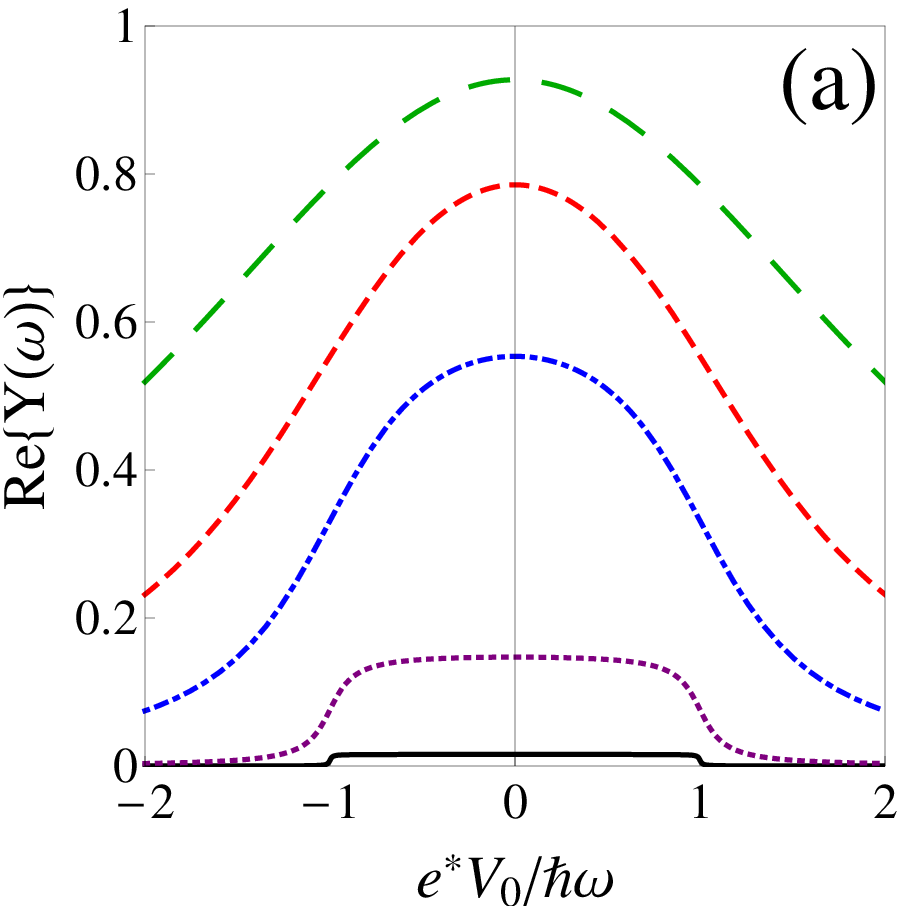}
\includegraphics[width=4.1cm]{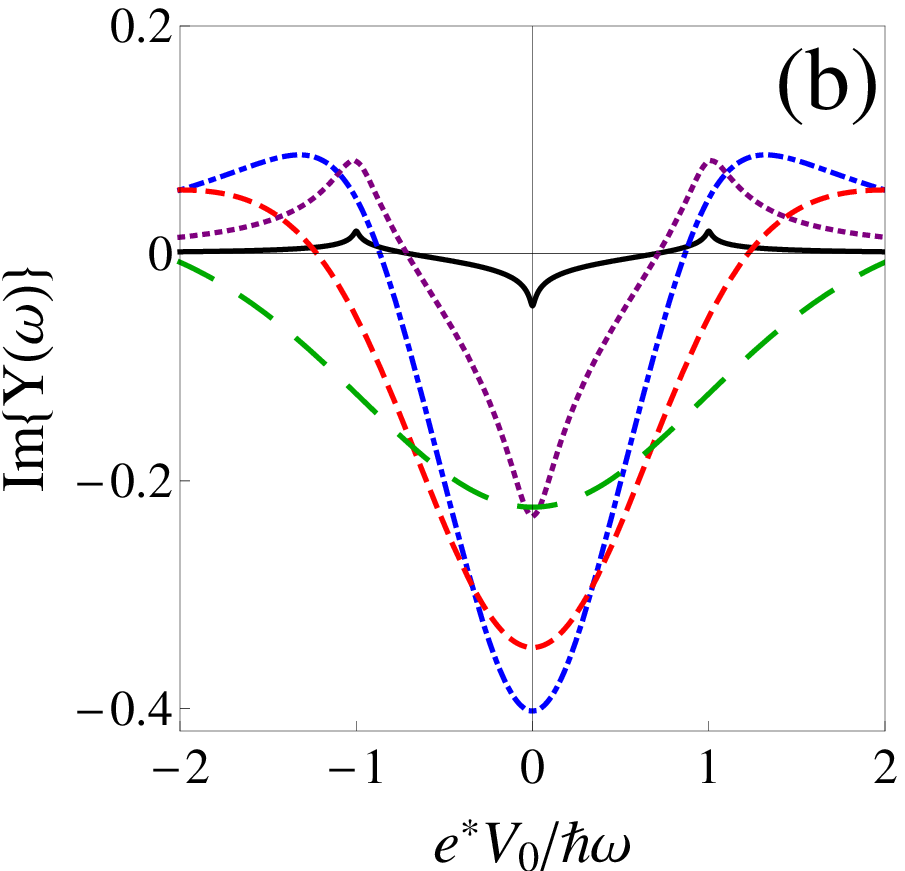}
\includegraphics[width=3.8cm]{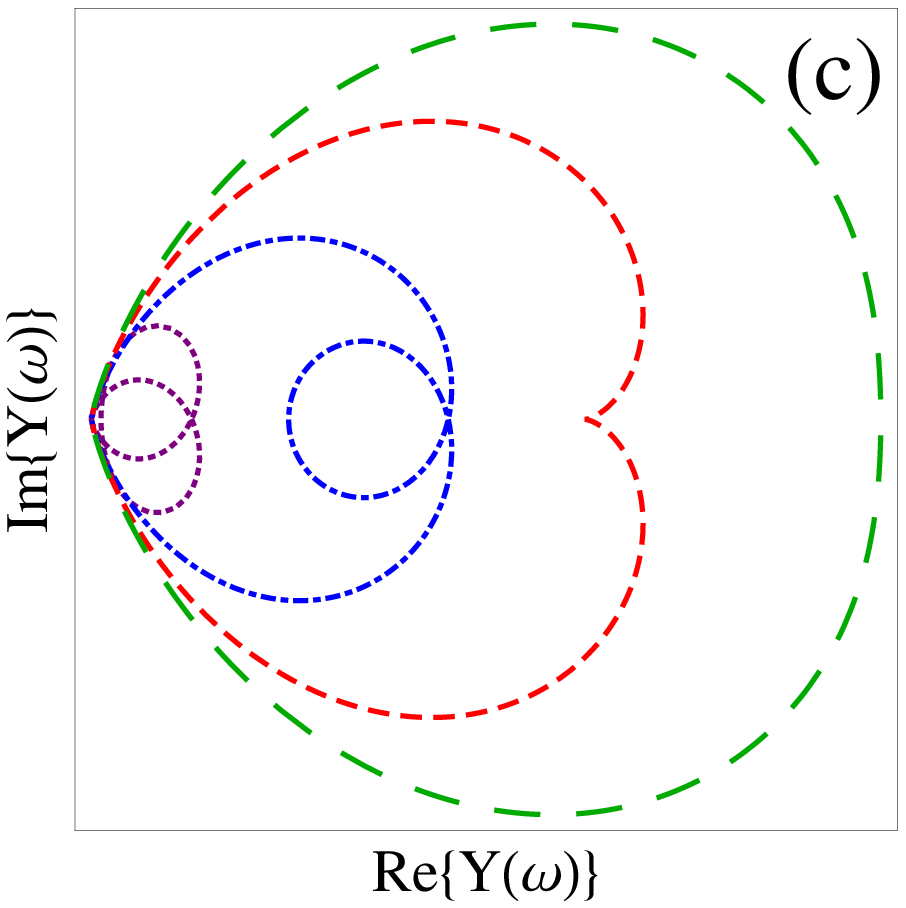}
\includegraphics[width=4cm]{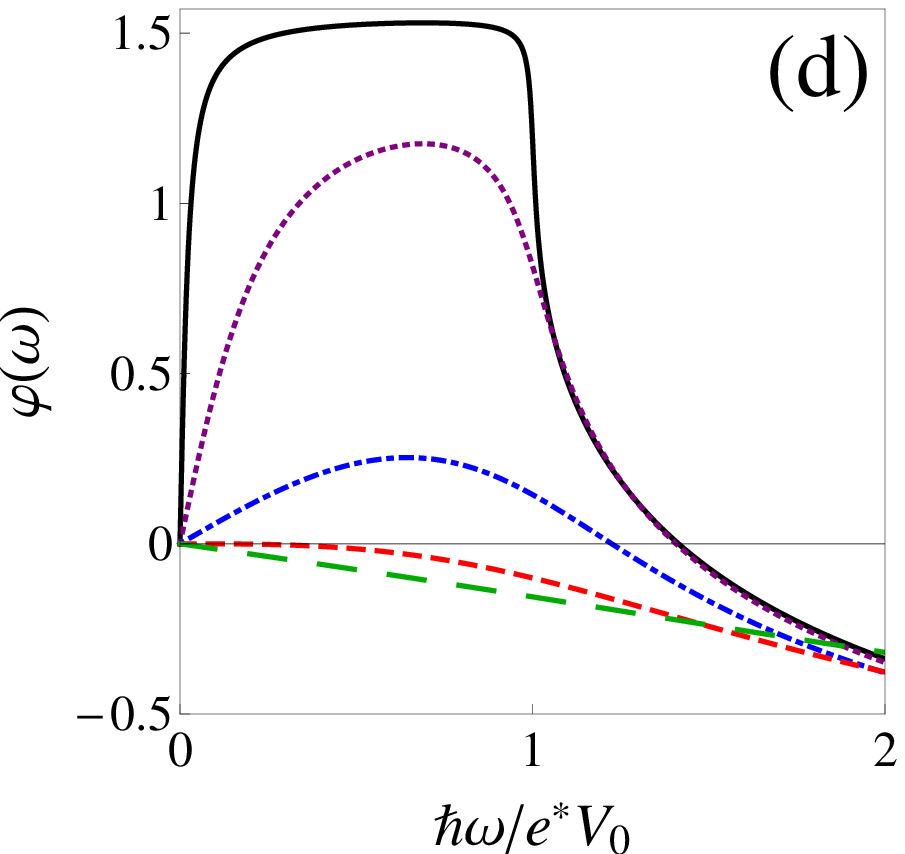}
\caption{(a)~Real part, (b)~imaginary part, (c)~Nyquist diagram, and (d) phase of the admittance at $T=0$, for $\hbar\omega_M=0.02e^*V_0$ (black solid line), $\hbar\omega_M=0.2e^*V_0$ (purple dotted line), $\hbar\omega_M=e^*V_0$ (blue dash-dotted line), $\hbar\omega_M=2e^*V_0$ (red dashed line) and  $\hbar\omega_M=4e^*V_0$ (green long dashed line). The black solid line in graph (c) is not shown for visibility reasons. $\mathrm{Re}\{Y(\omega)\}$ and $\mathrm{Im}\{Y(\omega)\}$ are depicted in units of $G_q=e^2/h$, the quantum of conductance.
\label{figure2}}
\end{center} 
\end{figure}

At zero temperature, the associated coefficients of the series expansion read as:
\begin{eqnarray}\label{series_T0}
Y_{T=0}^{(n)}&=&\frac{e^2}{4h}\sum_\pm\frac{(-i)^n\hbar^{n+1}\omega_M}{(n+1)(\hbar\omega_M/2\pm ie^*V_0)^{n+1}}~.
\end{eqnarray}
In Fig.~\ref{figure2}(a) is plotted the real part of the admittance at zero temperature as a function of voltage. Steps are observed at positions $e^*V_0=\pm\hbar\omega$ (see black solid and purple dotted lines) which disappear when $\hbar\omega_M$ increases. The profile of the imaginary part of the admittance shown in Fig.~\ref{figure2}(b) is in agreement with recent experimental data\cite{basset12} confirming that the distance between the two peaks observed at weak $\hbar\omega_M$ is equal to $2\hbar\omega$. Fig.~\ref{figure2}(c) reveals interesting features which highlight the fact that the regimes of weak and strong $\hbar\omega_M$ differ fundamentally. Indeed, in the limit $\hbar\omega_M\gg e^*V_0$, the Nyquist diagram is a quasi-circle (see the green long dashed line in Fig.~\ref{figure2}(c)) which means that the admittance is a function of a complex variable $z$ with a single pole $z_p$ of order 1, i.e. of the form: $(z-z_p)^{-1}$. When $\hbar\omega_M$ decreases, singularities and loops appear (see red dashed, blue dash-dotted and purple dotted lines in Fig.~\ref{figure2}(c)). This means that the admittance is a complex function with a single pole of higher order. As a consequence, the sign of the phase of the admittance will change when $\omega$ varies. It is precisely what is observed in Fig.~\ref{figure2}(d) where the phase of the admittance is depicted as a function of frequency. The sign of the phase gives indication about the inductive (i.e. $\varphi(\omega)<0$) or capacitive (i.e. $\varphi(\omega)>0$) character of the mesoscopic system. For weak values of $\hbar\omega_M$ and for frequency within the interval $[-e^*V_0/\hbar,e^*V_0/\hbar]$, the current is in phase opposition with the ac voltage since $\varphi(\omega)\approx\pi/2$ (see the black solid line) and the mesoscopic system behaves as a capacitor. At high frequency, the system becomes inductive whatever the value of $\hbar\omega_M$ is. This dependence of the phase with the frequency is in qualitative agreement with experimental data obtained recently in carbon nanotube.\cite{chorley12}

\subsection{Low frequency limit}

When the ac frequency $\omega$ is much smaller than all the other characteristic frequencies, i.e. $\omega_M$, $k_BT/\hbar$ and $e^*V_0/\hbar$, explicit expressions of the admittance up to the first order in $\omega$ can be obtained from Eq.~(\ref{coeff_n}) by retaining only the terms $n=0$ and $n=1$ in the sum over $n$. The results are summarized in the two first lines of Table~\ref{table1}. Again, it is interesting to interpret them in terms of an effective RLC circuit. At strong $\hbar\omega_M$, since $\varphi(\omega)<0$, the mesoscopic system is inductive whatever the temperature is: the mesoscopic system behaves as a RL circuit in series, with an effective resistance independent of any characteristic energy: $R_{\mathrm{eff}}=G_q^{-1}$, and an effective inductance which depends on $\omega_M$: $L_{\mathrm{eff}}=(\omega_MG_q)^{-1}$. Note that the admittance converges to $G_q$ at very large $\hbar\omega_M$.

\subsection{Other limits}

At weak $\hbar\omega_M$, the dynamics of the mesoscopic system is strongly temperature dependent: at low temperature (compared to the voltage), the system is capacitive ($\varphi(\omega)>0$) whereas it is inductive ($\varphi(\omega)<0$) at higher temperature. The fact that at zero temperature a change from capacitive to inductive circuit is observed when $\hbar\omega_M$ increases is in agreement with what was obtained in Ref.~\onlinecite{fu93} for a double barrier quantum dot. Indeed, at weak coupling $\hbar\omega_M\propto \Gamma_L+\Gamma_R$, this system is capacitive because of the small value of the transmission coefficient equal to $\mathcal{T}(\Omega)=t(\Omega)t^*(\Omega)$, with no possibility of activation by any energy since both temperature and frequencies are weak. It is interesting to notice that in the limit $\hbar\omega_M\approx 0$, the expression of the effective resistance, $R_\mathrm{eff}$, always contains the interactions/coupling energy, $\hbar\omega_M$, and the largest of the three other characteristic energies, i.e. either $k_BT$, $e^*V_0$, or $\hbar\omega$. Indeed, at high temperature: $G_qR_\mathrm{eff}=8k_BT/(\pi \hbar\omega_M)$,  in agreement with Ref.~\onlinecite{nigg06}; at high voltage: $G_qR_\mathrm{eff}=(2e^*V_0)^2/(\hbar\omega_M)^2$; and at high frequency: $G_qR_\mathrm{eff}=4\omega/(\pi\omega_M)$. Furthermore, at high temperature and whatever the value of $\hbar\omega_M$ is, the effective resistivity reads as:
\begin{eqnarray}
R_\mathrm{eff}^{-1}=\frac{\partial I^\mathrm{dc}(V_0)}{\partial V_0}=\frac{e^2\hbar\omega_M}{16\pi hk_BT}\Psi'\left(\frac{1}{2}+\frac{\hbar\omega_M}{8\pi k_BT}\right)~,
\end{eqnarray}
in agreement with Ref.~\onlinecite{fendley95} ($\Psi'$ is the derivative of the digamma function).

\begin{widetext}

\begin{table}[!h]
\begin{center}
\begin{tabular}{|l|l|l|}
\hline
& $\hbar\omega_M\approx 0$ & $\hbar\omega_M\gg\{k_BT,|e^*V_0|,|\hbar\omega|\}$\\ \hline
\multirow{3}{*}{$k_BT\ll |e^*V_0|$ and $\hbar\omega\approx 0$} & $Y(\omega)=G_q\left(\frac{\hbar\omega_M}{2e^*V_0}\right)^2\left(1+i\frac{\omega}{\omega_M}\right)$ & \\
& $\Rightarrow$ RC circuit in parallel &  \\
& $R_\mathrm{eff}^{-1}=G_q\left(\frac{\hbar\omega_M}{2e^*V_0}\right)^2$, $C_\mathrm{eff}^{-1}=\omega_MR_\mathrm{eff}$ &\\\cline{1-2}
\multirow{3}{*}{$k_BT\gg |e^*V_0|$ and $\hbar\omega\approx 0$} & $Y(\omega)=G_q\frac{\pi \hbar\omega_M}{8k_BT}\left(1-i\frac{\hbar\omega}{4k_B T}\right)$ & $Y(\omega)=G_q\left(1-i\frac{\omega}{\omega_M}\right)$\\
& $\Rightarrow$ RL circuit in series & $\Rightarrow$ RL circuit in series \\
& $R_\mathrm{eff}^{-1}=G_q\frac{\pi \hbar\omega_M}{8k_BT}$, $L_\mathrm{eff}^{-1}=\pi\omega_MG_q/2$& $R_{\mathrm{eff}}^{-1}=G_q$, $L_{\mathrm{eff}}^{-1}=\omega_MG_q$ \\ \cline{1-2}
& $Y(\omega)=G_q\frac{\omega_M}{4\omega}\left[\pi\mathrm{sign}(\omega)-i\ln\left(\frac{(\hbar\omega)^2}{(\hbar\omega_M/2)^2+(e^*V_0)^2}\right)\right]$ &\\ 
$k_BT\ll |e^*V_0|\ll|\hbar\omega|$&$\Rightarrow$ RL circuit in parallel &\\
&$R_\mathrm{eff}^{-1}=G_q\frac{\pi\omega_M}{4|\omega|}$,$L_\mathrm{eff}^{-1}=\frac{G_q\omega_M}{4}\ln\left(\frac{(\hbar\omega)^2}{(\hbar\omega_M/2)^2+(e^*V_0)^2}\right)$&\\
\hline
\end{tabular}
\caption{Admittance $Y(\omega)$ and its associated effective resistance and inductance/capacitance in various limits. $G_q=e^2/h$ refers to the quantum of conductance.}
\label{table1}
\end{center}
\end{table}

\end{widetext}

\begin{figure}[!h]
\begin{center}
\includegraphics[width=4.2cm]{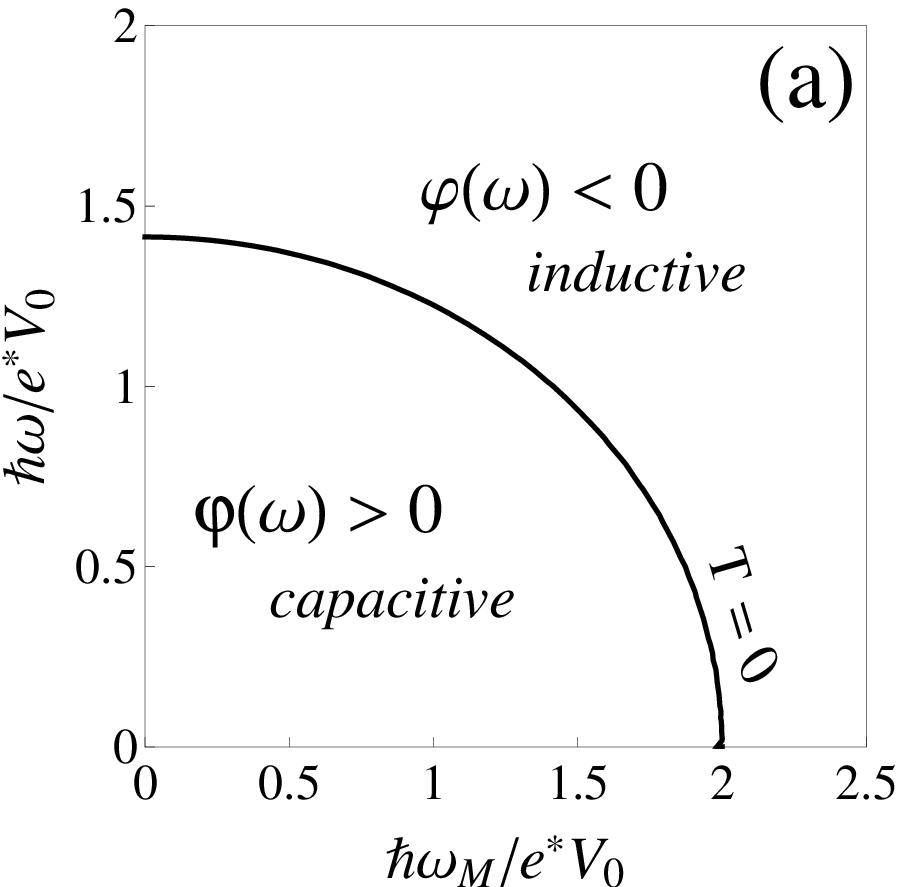}
\includegraphics[width=4.2cm]{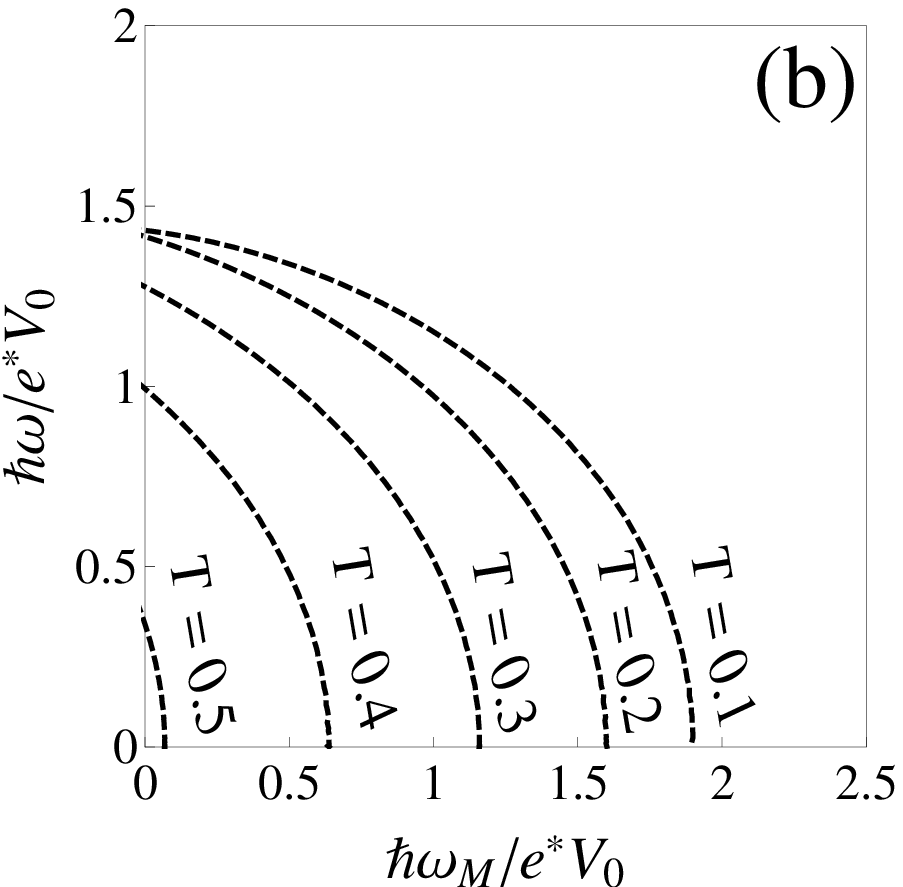}
\caption{(a) Sign of the phase of the admittance at zero temperature as a function of coupling/interaction $\hbar\omega_M$ and  ac frequency $\hbar\omega$. (b) Evolution of the boundary between regions of positive and negative signs when temperature $T$ increases (in units of $e^*V_0/k_B$).
\label{figure3}}
\end{center} 
\end{figure}

In order to understand the relative effects of the various energies that characterize the system, the boundary between regions of positive and negative phases is drawn in Fig.~\ref{figure3} as a function of $\hbar\omega_M$ and $\hbar\omega$, both taken in units of $e^*V_0$. At zero temperature, the equation for the boundary line can be extracted from Eq.~(\ref{Y_T0}), it reads as: $|\hbar\omega|=\sqrt{2[(e^*V_0)^2-(\hbar\omega_M/2)^2]}$. Below this line, the phase is positive whereas above it, the phase is negative (see Fig.~\ref{figure3}(a)). The phase separation line moves to the origin when the temperature increases and so the region of positive phase collapses (see Fig.~\ref{figure3}(b)). Accordingly, to get a positive value for the phase, i.e. a capacitive behavior of the mesoscopic system, both temperature and characteristic frequencies must be small enough in comparison to the dc voltage. More precisely, one must have: $\hbar\omega_M/2\lesssim|e^*V_0|$, $|\hbar\omega|\lesssim\sqrt{2[(e^*V_0)^2-(\hbar\omega_M/2)^2]}$, and $k_BT\lesssim|e^*V_0|$.

\section{Conclusion}

The quantum admittance associated to one impurity in a Tomonaga-Luttinger liquid with $\nu=1/2$ has been calculated with the help of a refermionization method. Its expression is identical to the one of a double barrier quantum dot. Furthermore, in light of the mapping of Ref.~\onlinecite{safi04}, it is possible to consider in an unified view the three systems depicted in Fig.~\ref{figure1} and to give an unique and compact expression for their admittance. A detailed analysis of the various limits has shown that if at least one of the characteristic energies is larger than the dc applied voltage, the mesoscopic system behaves has an inductance since it is activated by either temperature, frequency or interactions/coupling frequency.

\section{Acknowledgement}

The author thanks M. Lavagna, I. Safi and R. Zamoum for discussions.

\appendix 

\section{Admittance calculation for system (c) depicted on Fig.~\ref{figure1}}\label{appA}

The time-dependent electric current which flows from the left (L) or the right (R) reservoir to the central region of a double barrier junction reads as:\cite{jauho94}
\begin{eqnarray}
&&I_{L,R}(t)=\frac{2e\Gamma_{L,R}}{2\pi}\int_{-\infty}^{\infty} f(\hbar\Omega-eV_{L,R})\mathrm{Im}\{A(\Omega,t)\} d\Omega~.\nonumber\\
\end{eqnarray}
In the case where an ac voltage superposed to a dc voltage $V(t)=V_0+V_\omega\cos(\omega t)$ is applied to the dot, the spectral function $A$ is given by:
\begin{eqnarray}
A(\Omega,t)&=&\sum_{p=-\infty}^{\infty}\sum_{m=-\infty}^{\infty}J_p\left(\frac{eV_\omega}{\hbar\omega}\right)J_m\left(\frac{eV_\omega}{\hbar\omega}\right)\nonumber\\
&&\times\frac{e^{i(p-m)\omega t}}{E_p(\Omega)+i\hbar\omega_M/2}~,
\end{eqnarray}
where $\hbar\omega_M=\Gamma_L+\Gamma_R$, $E_p(\Omega)=\hbar\Omega-eV_0-p\hbar\omega$, and $J_p$ is the Bessel function of order $p$.  From this expression, the harmonics of the photo-assisted current can be extracted:
\begin{eqnarray}
I_{L,R}^{(N)}&=&\frac{ie\Gamma_{L,R}}{2\pi}\sum_{\pm}\Bigg\{\pm\sum_{p=-\infty}^{\infty}J_p\left(\frac{eV_\omega}{\hbar\omega}\right)J_{p\pm N}\left(\frac{eV_\omega}{\hbar\omega}\right)\nonumber\\
&&\times\int_{-\infty}^{\infty}\frac{f(\hbar\Omega-eV_{L,R})}{E_p(\Omega)\mp i\hbar\omega_M/2}d\Omega\Bigg\}~.
\end{eqnarray}

By considering that the left and right reservoirs are set at the same voltage, i.e. $V_L=V_R=0$, the first harmonic of the current reduces to:
\begin{eqnarray}\label{first_harmonic}
I_{L,R}^{(1)}&=&\frac{ie\Gamma_{L,R}}{2\pi}\sum_{\pm}\Bigg\{\pm\sum_{p=-\infty}^{\infty}J_p\left(\frac{eV_\omega}{\hbar\omega}\right)J_{p\pm 1}\left(\frac{eV_\omega}{\hbar\omega}\right)\nonumber\\
&&\times\int_{-\infty}^{\infty}\frac{f(\hbar\Omega)}{E_p(\Omega)\mp i\hbar\omega_M/2}d\Omega\Bigg\}~.
\end{eqnarray}

Using the fact that in the limit $x\rightarrow 0$, we have $J_0(x)\approx 1$ and $J_{\pm 1}(x)\approx \pm x/2$, the first harmonic of the current up to the first order in $V_\omega$ (linear response) reduces to:
\begin{eqnarray}
I_{L,R}^{(1)}&=&\frac{ie\Gamma_{L,R}}{2\pi}\sum_{\pm}\left(\frac{eV_\omega}{2\hbar\omega}\right)\int_{-\infty}^{\infty}\frac{ f(\hbar\Omega)}{E_0(\Omega)\mp i\hbar\omega_M/2}d\Omega\nonumber\\
&&-\frac{ie\Gamma_{L,R}}{2\pi}\left(\frac{eV_\omega}{2\hbar\omega}\right)\int_{-\infty}^{\infty}\frac{ f(\hbar\Omega)}{E_1(\Omega)+ i\hbar\omega_M/2}d\Omega\nonumber\\
&&-\frac{ie\Gamma_{L,R}}{2\pi}\left(\frac{eV_\omega}{2\hbar\omega}\right)\int_{-\infty}^{\infty}\frac{ f(\hbar\Omega)}{E_{-1}(\Omega)- i\hbar\omega_M/2}d\Omega~.\nonumber\\
\end{eqnarray}

The first, second and third terms in this expression correspond respectively to the terms $p=0$, $p=1$ and $p=-1$ in the sum over $p$ of Eq. (\ref{first_harmonic}). Finally:
\begin{eqnarray}
I^{(1)}(\omega)&=&\frac{I_{L}^{(1)}+I_{R}^{(1)}}{2}=\frac{e^2V_\omega}{2h\omega}\int_{-\infty}^{\infty}f(\hbar\Omega)\nonumber\\
&&\times\Big[t(\Omega-eV_0/\hbar)-t(\Omega-\omega-eV_0/\hbar)\nonumber\\
&&-t^*(\Omega-eV_0/\hbar)+t^*(\Omega+\omega-eV_0/\hbar)\Big]d\Omega~,\nonumber\\
\end{eqnarray}
which leads after a change of variables to the admittance:
\begin{eqnarray}\label{admittance_system_c}
Y(\omega)&=&\frac{\partial I^{(1)}(\omega)}{\partial V_\omega}=\frac{e^2}{2h\omega}\int_{-\infty}^{\infty}f(\hbar\Omega+eV_0)\nonumber\\
&&\times\Big[t(\Omega)-t(\Omega-\omega)-t^*(\Omega)+t^*(\Omega+\omega)\Big]d\Omega~.\nonumber\\
\end{eqnarray}
This result is identical to Eq.~(\ref{exp_admi}) since $e^*=e$ for the system (c) of Fig. 1.

\section{Series expansion of the admittance}\label{appB}

Eq.~(\ref{exp_admi}) can be expanded in powers of $\omega$. The starting point is the definition of the amplitude $t(\Omega-\omega)$ which appears in the expression of the admittance that can be transformed in this way:
\begin{eqnarray}
t(\Omega-\omega)&=&\frac{i\omega_M/2}{\Omega-\omega+i\omega_M/2}\nonumber\\
&=&\frac{i\omega_M/2}{(\Omega+i\omega_M/2)\left(1-\frac{\omega}{\Omega+i\omega_M/2}\right)}~.
\end{eqnarray}
It can be expressed in term of amplitude $t(\Omega)$ through:
\begin{eqnarray}
t(\Omega-\omega)&=&t(\Omega)\sum_{n=0}^\infty\left(\frac{\omega}{\Omega+i\omega_M/2}\right)^n\nonumber\\
&=&t(\Omega)\sum_{n=0}^\infty\left(\frac{\omega}{i\omega_M/2}\right)^n\left(\frac{i\omega_M/2}{\Omega+i\omega_M/2}\right)^n\nonumber\\
&=&t(\Omega)\left(1+\sum_{n=1}^\infty\left(\frac{\omega}{i\omega_M/2}\right)^n\left[t(\Omega)\right]^n\right)~.\nonumber\\
\end{eqnarray}

Reporting this result in the expression~(\ref{exp_admi}) of the admittance, it leads to:
\begin{eqnarray}
Y(\omega)&=&
-\frac{e^2}{2h}\sum_{n=0}^\infty\frac{\omega^{n}}{(i\omega_M/2)^{n+1}}\int_{-\infty}^{\infty}\left[t(\Omega)\right]^{n+2}\nonumber\\
&&\times\big[f(\hbar\Omega+e^*V_0)-f(-\hbar\Omega+e^*V_0)\big]d\Omega~,\nonumber\\
\end{eqnarray}
which finally gives the expression~(\ref{coeff_n}) of the harmonics of the admittance.



\begin{thebibliography}{20}
\bibitem{note1}The following convention has been used: $Y(\omega)$ is the coefficient in front of the $e^{i\omega t}$ term in the electric current response $I(t)$. This convention must be kept in mind when one compares results of this work to the literature.
\bibitem{note2}Strictly speaking, Eq.~(\ref{hamiltonian}) describes incompressible chiral edge modes and thus is not applicable to $\nu=1/2$ states which are expected to be compressible. On the opposite, the refermionization procedure is only possible at $\nu=1/2$. Despite this discrepancy, it is believed that the refermionization procedure which enables exact calculations can provide valuable information.
\end{thebibliography}
\end{document}